\documentclass[showpacs,preprintnumbers,pre,twocolumn]{revtex4}
\usepackage{graphicx}
\usepackage{dcolumn}
\usepackage{bm}

\input{tcilatex}

\begin{document}

\preprint{}
\title{Influence of Correlated Noises on Growth of a Tumor}
\author{Wei-Rong Zhong}
\author{Yuan-Zhi Shao}
\altaffiliation[ ]{Corresponding Author}
\email{stssyz@zsu.edu.cn}
\author{Zhen-Hui He}
\affiliation{Department of Physics, Sun Yat-sen University, 510275 Guangzhou, People's
Republic of China}

\begin{abstract}
We studied the effect of additive and multiplicative noises on the growth of
a tumor based on a logistic growth model. The steady-state probability
distribution and the average population of the tumor cells were given to
explain the important roles of correlated noises in the tumor growth. We
found that multiplicative noise induces a phase transition of the tumor
growth from an uni-stable state to a bi-stable state; the relationship
between the intensity of multiplicative noise and the population of the
tumor cells showed a stochastic resonance-like characteristic. It was also
confirmed that additive noise weakened rather than extinguish the tumor
growth. Homologous noises, however, promote the growth of a tumor. We also
discussed about the relationship between the tumor treatment and the model.
\end{abstract}

\maketitle

\begin{center}
\textbf{I. INTRODUCTION}
\end{center}

Nonlinear Physics has changed people's views of world drastically
in the past half a century [1, 2], and it is expected useful for
discovering the complexity of Biology [3]. More and more attempts
confirmed the important role of noise in the nonlinear dynamic
systems [4, 5],. Especially, Fokker-Planck equation became one of
the approaches of nonlinear dynamics based on deterministic
equation with the stochastic form [6, 7]. During the last twenty
years, bistable-state model is one of those that attract
Physicists most [8-15].

For many years, exploring the growth law of tumors is a challenging subject.
Mathematical models are always used as a tentative for describing the tumor
growth [16-26]. Bru et al [23, 24] discovered that tumor growth was largely
governed by environmental pressures (host tissue pressure, immune response,
etc.) through both simulations and experiments of the immune responses of
mice. Chemotherapeutic double-faces effect on the cancer was studied\ by
Ferreira et al [25]. It was confirmed that the environment like chemotherapy
affected the growth of cancer vitally. Recently, Ai et al [26] applied
Logistic growth model to analyze the growth of tumors driven by correlated
noises. The relationship between correlated noises and tumor growth was
mentioned for the first time. For simplicity, Ai et al only investigated the
influence of multiplicative noise on the growth rate of tumors. They found
that additive and multiplicative noises would lead the tumors to extinction.
They, however, only performed a qualitative analysis according to the
steady-state probability distribution of tumor cells. It is unclear whether
the influence of correlated noises on a tumor is always negative.

Tumor growth is a complex process. The search based on tumor growth dynamics
for the underlying mechanisms of the tumor development and progression might
focus on lots of factors. On one hand, the environmental factors, like
temperature, chemotherapy and radiotherapy, may affect the decay rate as
well as the growth rate; on the other hand, intrinsic elements, like
emigration and genetic instability, also will influence the growth of tumor
cells strongly [27]. The environmental factors were regarded as the sources
of multiplicative noise, while the intrinsic elements as the additive ones.
Considering the fluctuation of the linear growth rate under a stable
carrying capacity of environment, we studied the influence of the
intensities of correlated noises on the growth and decay of tumor cells.
Stochastic resonance (SR) is another concern of this paper. Generally,
bistable state is one of the conditions for SR [4]. However, the Logistic
model, a non-bistable state system driven by correlated noises as showed in
this paper, also shows a typical SR-like characteristic. It is of interest
to study this phenomenon as well as its underlying mechanism. We attempt to
bridge the gap between correlated noises and tumor treatment and give a new,
helpful concept for the treatment of cancer.

\begin{center}
\textbf{II. LANGEVIN EQUATION AND FOKKER-PLANCK EQUATION}
\end{center}

Verhulst [28] proposed a single-species model with a self-limiting process.
He suggested%
\begin{equation}
\frac{dx}{dt}=rx(1-\frac{x}{K})  \tag{1}
\end{equation}

This is the Logistic growth equation, where $x$ is the population; $r$ and $%
K $ are the linear per capita birth rate and the carrying capacity of the
environment, respectively. Eq.(1) was used to study the growth of the cells.
When the fluctuations of the environment, such as radiotherapy and
chemotherapy, were considered, the linear internal growth rate of tumor
cell, $r$, would not be a constant any more. Therefore, it is reasonable to
introduce noise $\xi (t)$ into the single-species model, and $r$ should be
rewritten as $r+\xi (t)$. Likewise, intrinsic factors were set as another
noise $\eta (t)$, e.g. emigration, selection pressure and genetic
instability, which could affect the population of tumor cells in local area.
So if $x$ was defined as the population of tumor cells, the Langevin
differential equation driven by correlated noises could be written as%
\begin{equation}
\frac{dx}{dt}=rx(1-\frac{x}{K})+x(1-\frac{x}{K})\xi (t)+\eta (t)  \tag{2}
\end{equation}%
\ where \ $\xi (t)$, $\eta (t)$ are respectively the multiplicative and the
additive Gaussian white noises defined as%
\begin{equation}
\langle \xi (t)\rangle =0,\ \ \langle \xi (t)\xi (t^{\prime })\rangle
=2D\delta (t-t^{\prime })  \tag{3a}
\end{equation}%
\begin{equation}
\langle \eta (t)\rangle =0,\ \ \langle \eta (t)\eta (t^{\prime })\rangle
=2Q\delta (t-t^{\prime })  \tag{3b}
\end{equation}%
\begin{equation}
\langle \xi (t)\eta (t^{\prime })\rangle =2\lambda \sqrt{DQ}\delta
(t-t^{\prime })  \tag{3c}
\end{equation}%
\ \ in which $D$ and $Q$ are the intensities of the noises, and $\lambda $
is the strength of the correlation between $\xi (t)$ and $\eta (t)$
respectively, and $\delta (t-t^{\prime })$ is Dirac delta function in
different times.

According to the Langevin equation (2), one would write out the equivalent
Fokker-Planck equation [29].%
\begin{equation}
\frac{\partial p(x,t)}{\partial t}=-\frac{\partial }{\partial x}[A(x)p(x,t)]+%
\frac{\partial ^{2}}{\partial x^{2}}[B(x)p(x,t)]  \tag{4}
\end{equation}%
where $p(x,t)$ is the probability distribution function, $A(x)$ and $B(x)$
are respectively defined as%
\begin{equation}
A(x)=rx(1-\frac{x}{K})+Dx(1-\frac{x}{K})(1-\frac{2x}{K})+\lambda \sqrt{DQ}(1-%
\frac{2x}{K})  \tag{5a}
\end{equation}%
\begin{equation}
B(x)=Dx^{2}(1-\frac{x}{K})^{2}+2\lambda \sqrt{DQ}x(1-\frac{x}{K})+Q  \tag{5b}
\end{equation}

The population of the cells, $x,$ is positive. According to the reflecting
boundary condition, the steady-state probability distribution (STPD) of
Eq.(4) is [9]%
\begin{equation}
p_{st}(x)=\frac{N}{B(x)}\exp [\int^{x}\frac{A(x^{^{\prime }})}{B(x^{^{\prime
}})}dx^{^{\prime }}]  \tag{6}
\end{equation}%
where $N$ is a normalization constant. From this equation, we attained the
STPD of the cells population. The peak position of the STPD is $K$ when $%
\lambda $ equals zero.

The following parameters are defined to analyze the decay and growth of
tumor cells quantitatively.%
\begin{equation}
\langle x\rangle =\frac{\int_{0}^{\infty }x\text{ }p_{st}(x)dx}{%
\int_{0}^{\infty }p_{st}(x)dx}  \tag{7a}
\end{equation}%
\begin{equation}
\langle x^{2}\rangle =\frac{\int_{0}^{\infty }x^{2}\text{ }p_{st}(x)dx}{%
\int_{0}^{\infty }p_{st}(x)dx}  \tag{7b}
\end{equation}%
\begin{equation}
\ \langle \delta x^{2}\rangle =\langle x^{2}\rangle -\langle x\rangle ^{2}
\tag{7c}
\end{equation}%
\ \ in which $\langle x\rangle ${\normalsize \ and }$\langle \delta
x^{2}\rangle ${\normalsize \ refer to the average population of tumor cells
and its square difference, respectively. Increasing }$\langle x\rangle $%
{\normalsize \ means the growing of the tumor cells, otherwise the decaying.
Augment of }$\langle \delta x^{2}\rangle $ denotes that the tumor growth is
unstable, otherwise\ it is stable.

\begin{center}
\textbf{III. ANALYSIS OF THE GROWTH MODEL OF A TUMOR }
\end{center}

Whether a tumor come into being depends on the linear growth rate, $r$, in
Eq.(1). If $r$ and $K$ are positive, e.g. $0.2\ $and $6$ respectively$,$ the
tumors will grow; else if $r$ and $K$ are negative, e.g. $-0.2,$ $-6$, the
tumors will decay. The noises induce different responses of tumor cells
between the growth and decay case.

\begin{center}
\textbf{A.\qquad Growth Case}
\end{center}

The influence of additive noise on the growth of a tumor is shown in Fig.1.
In the absence of noises, $p_{st}(x)=\delta (x-K)$, tumors population is
stable at $x=K$. In the presence of noises, a STPD appears with its peak at
the position $x=K$ (see Fig.1). As the intensity of the noise increases, the
peak of STPD decreases. When $D<0.2,$the population of tumor cells decreases
firstly and then increases with the intensity of additive noise, showing a
negative stochastic resonance-like characteristic. When $D>0.2$, the
stochastic resonance-like characteristic extincts (see Inset in Fig.1).%
\FRAME{ftbpFU}{3.0415in}{2.3445in}{0pt}{\Qcb{The dependence of the STPD on
the tumor cells population, the parameters of noises are $\protect\lambda %
=0.0,$ and $D=0.5$ respectively. Inset: the average population varies with
the intensity of additive noise, from top to bottom, $D$ is $0.20,$ $0.08$
and $0.02,$ respectively.}}{}{graph1.eps}{\special{language "Scientific
Word";type "GRAPHIC";maintain-aspect-ratio TRUE;display "USEDEF";valid_file
"F";width 3.0415in;height 2.3445in;depth 0pt;original-width
4.3033in;original-height 3.3105in;cropleft "0";croptop "1";cropright
"1";cropbottom "0";filename '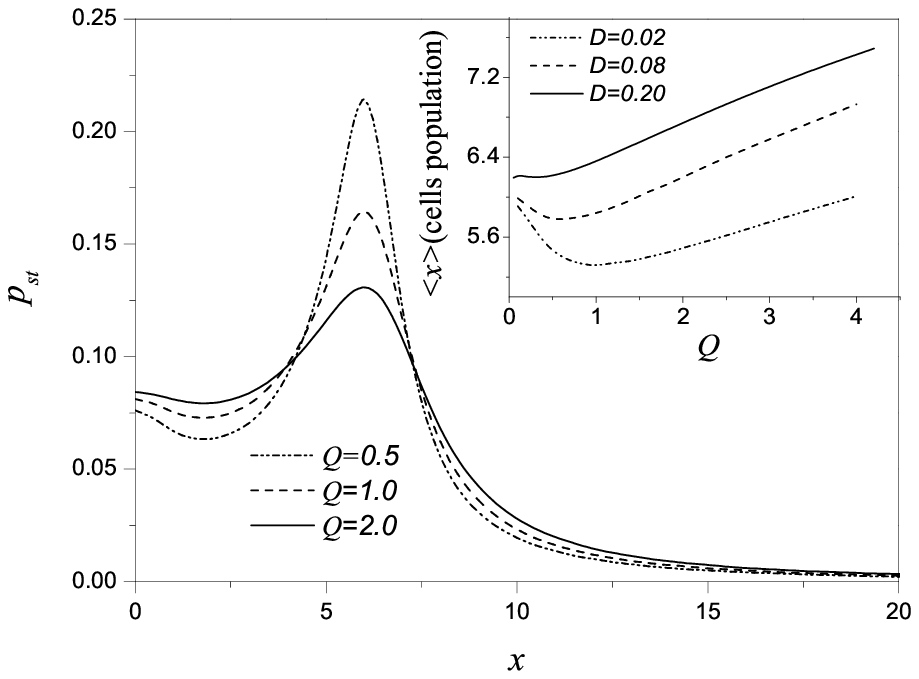';file-properties "XNPEU";}}

In addition to that of the additive noise, the influence of noises
correlation on the tumor growth was also studied. The STPD, as shown in
Fig.2, increases with the correlation parameter, $\lambda $. This implies
that the homologous noises promote the stable growth of tumor cells. Inset
shows increasing $\lambda $ is of benefit to tumor growth in the case of low
noise intensity; contrarily, increasing $\lambda $ is unfavorable to the
survival of tumor cells. $\lambda $ has little effect on the tumors at
intermediate intensity.\FRAME{ftbpFU}{2.757in}{2.1248in}{0pt}{\Qcb{STPD
under different correlations of noises, the noise parameters are $D=0.5$ and
$Q=2.0.$ Inset: the average population varies with the strength of
correlation.}}{}{graph2.eps}{\special{language "Scientific Word";type
"GRAPHIC";maintain-aspect-ratio TRUE;display "USEDEF";valid_file "F";width
2.757in;height 2.1248in;depth 0pt;original-width 4.3033in;original-height
3.3105in;cropleft "0";croptop "1";cropright "1";cropbottom "0";filename
'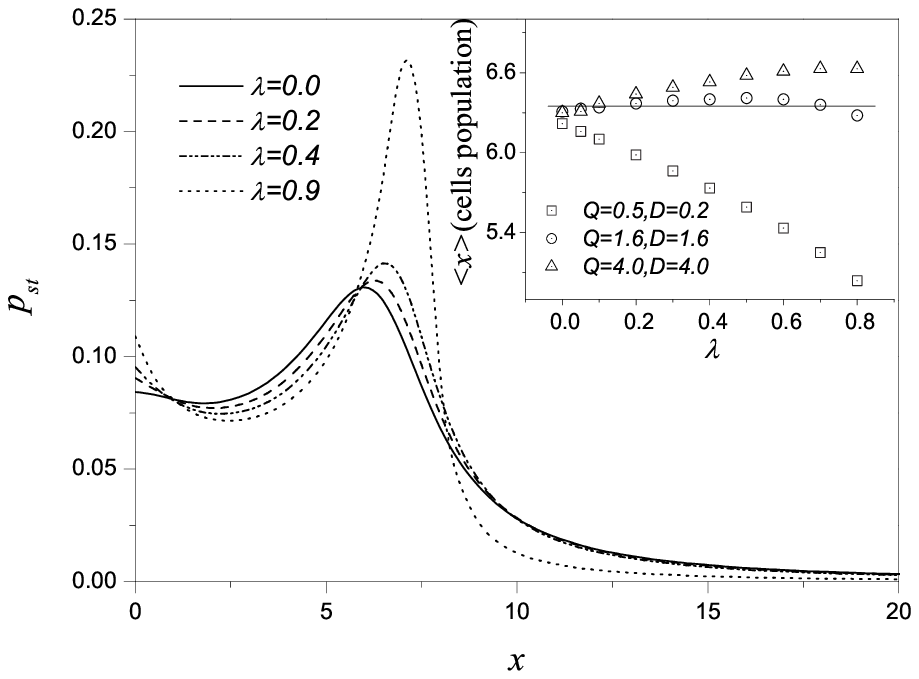';file-properties "XNPEU";}}

Unlike the additive noise, multiplicative noise has a interesting influence
on tumor growth. Fig.3 shows a STPD transition from an uni-peak to a
bi-peak\ distribution with the noise intensity. Moreover, the stochastic
resonance induced by multiplicative noise is also given. In Fig.4, $\langle
x\rangle $ increases firstly and then decreases with the intensity of
multiplicative noise, showing a typical stochastic resonance characteristic.
This means an appropriate intensity of multiplicative noise is suitable for
tumor cells, and extra noise restrains tumor growth, even leads them to
extinction. Fig.5 shows the influence of noises correlation on the
stochastic resonance induced by multiplicative noise. When $\lambda >0.8$,
the stochastic resonance peak will degenerate and disappear. Inset of Fig.5
shows the position $D_{peak}$, the peak of $\langle x\rangle $, increases
exponentially with the parameter, $\lambda $.

\FRAME{ftbpFU}{3.1254in}{2.4085in}{0pt}{\Qcb{STPD under different
multiplicative noise intensities, the parameters are $\protect\lambda =0.0$
and $Q=2.0$.}}{}{graph3.eps}{\special{language "Scientific Word";type
"GRAPHIC";maintain-aspect-ratio TRUE;display "USEDEF";valid_file "F";width
3.1254in;height 2.4085in;depth 0pt;original-width 4.3033in;original-height
3.3105in;cropleft "0";croptop "1";cropright "1";cropbottom "0";filename
'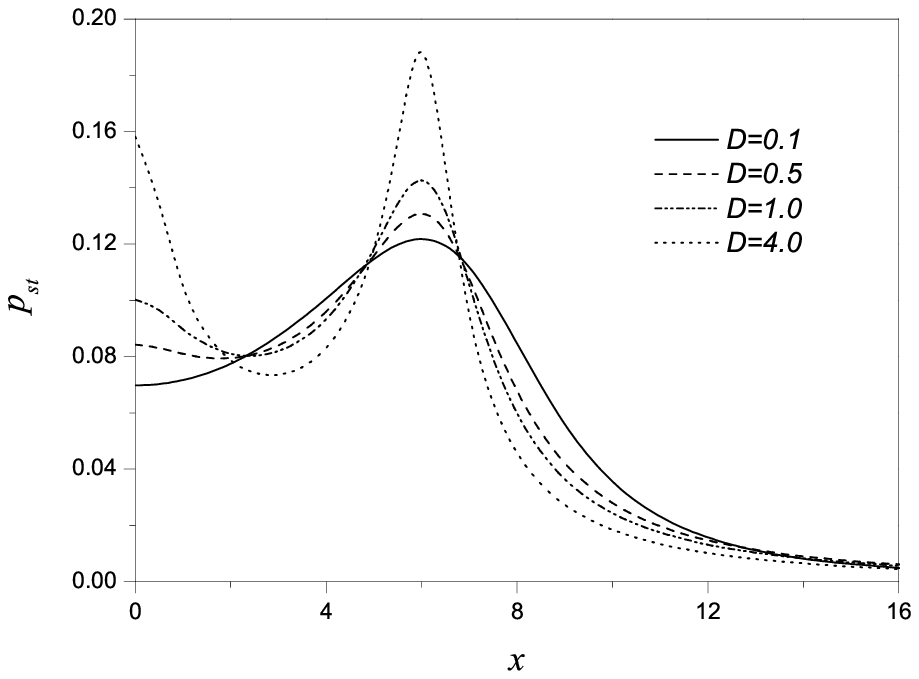';file-properties "XNPEU";}}\FRAME{ftbpFU}{3.1981in}{2.4414in}{0pt%
}{\Qcb{The change of the average population of cells against the intensity
of multiplicative noise at $\protect\lambda =0.0$, from bottom to top, $%
Q=0.5,1.0,2.0,$ respectively. }}{}{graph4.eps}{\special{language "Scientific
Word";type "GRAPHIC";maintain-aspect-ratio TRUE;display "USEDEF";valid_file
"F";width 3.1981in;height 2.4414in;depth 0pt;original-width
4.0932in;original-height 3.1176in;cropleft "0";croptop "1";cropright
"1";cropbottom "0";filename '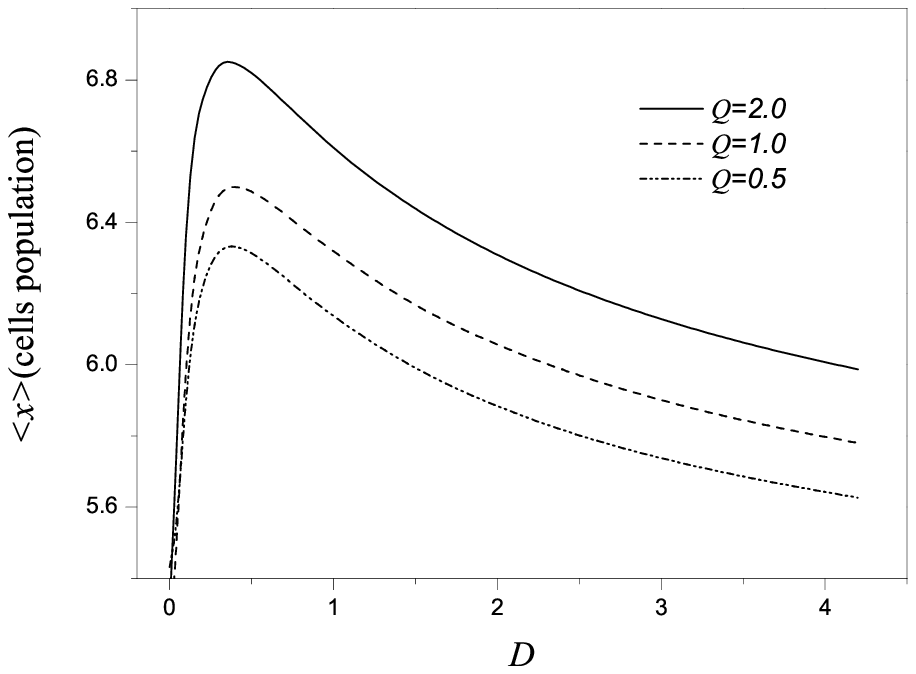';file-properties "XNPEU";}}\FRAME{%
ftbpFU}{3.2837in}{2.578in}{0pt}{\Qcb{The change of the average population of
cells with the intensity of multiplicative noise at $Q=2.0.$ Inset: the
position, $D_{peak},$ of the peak of the STPD changes with $\protect\lambda %
. $}}{}{graph5.eps}{\special{language "Scientific Word";type
"GRAPHIC";maintain-aspect-ratio TRUE;display "USEDEF";valid_file "F";width
3.2837in;height 2.578in;depth 0pt;original-width 4.2601in;original-height
3.3382in;cropleft "0";croptop "1";cropright "1";cropbottom "0";filename
'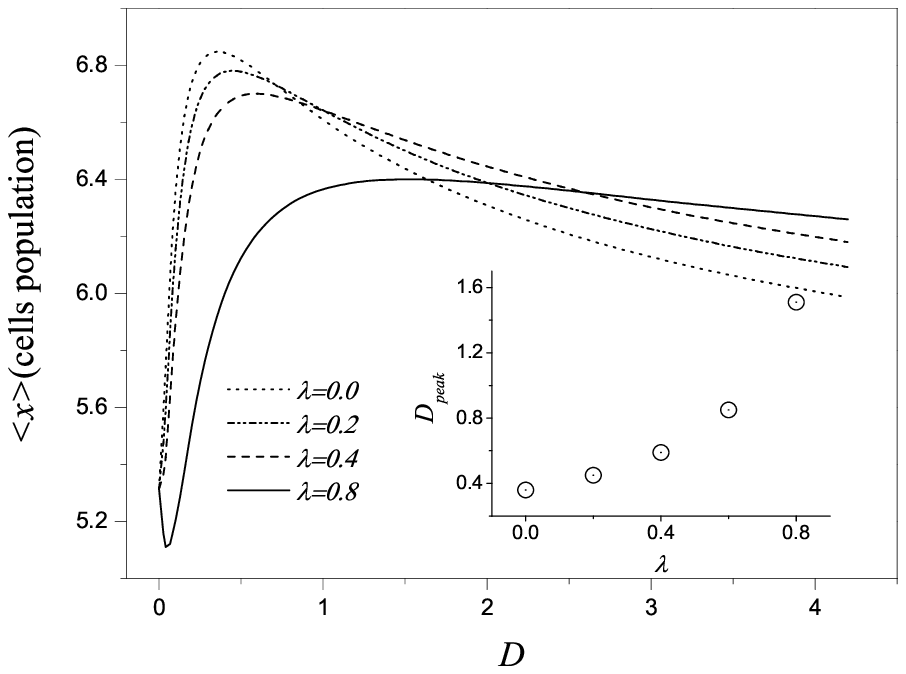';file-properties "XNPEU";}}

\begin{center}
\textbf{B.\qquad Decay Case}
\end{center}

In the decay case, the influence of noise intensity and correlation on the
tumors growth is shown in Fig.6-8. Similar to the growth case, in the
absence of noises, $p_{st}(x)=\delta (x-0)$, tumors could not grow. After
introducing noises, one could find that some tumor cells are still alive. In
Fig.6, the peak of STPD drops and $\langle \delta x^{2}\rangle $ increases
with the additive noise intensity. This is in agreement with the results of
Fig.1 that additive noise always causes an unstable growth of tumor cells.
Different from that of the growth case, the noises correlation has little
effect on the STPD of tumor growth in the decay case (see Fig.7). However,
inset of Fig.7 shows that the population of tumor cells still increase with
the strength of the noises correlation. This denotes that strong correlation
would keep the tumor far from extinction. In addition, just like that in the
growth case, multiplicative noise also induces a stochastic resonance in the
decay case. Fig.8 shows that the peak of STPD increases firstly and then
decrease with the intensity of multiplicative noise. There is shown in the
inset that the average population of cells, $\langle x\rangle ,$ and the
peak of STPD at $x=0,$ $p_{st}(0),$ change with $D.$ It is also a stochastic
resonance-like characteristic in the decay case (see the inset of Fig.8).%
\FRAME{ftbpFU}{2.9654in}{2.316in}{0pt}{\Qcb{The STPD of tumor cells
population changes with the intensity of additive noise at $\protect\lambda %
=0.0,D=0.5.$ Inset: the varieties of $\langle \protect\delta x^{2}\rangle $
with $Q.$ }}{}{graph6.eps}{\special{language "Scientific Word";type
"GRAPHIC";maintain-aspect-ratio TRUE;display "USEDEF";valid_file "F";width
2.9654in;height 2.316in;depth 0pt;original-width 4.248in;original-height
3.3105in;cropleft "0";croptop "1";cropright "1";cropbottom "0";filename
'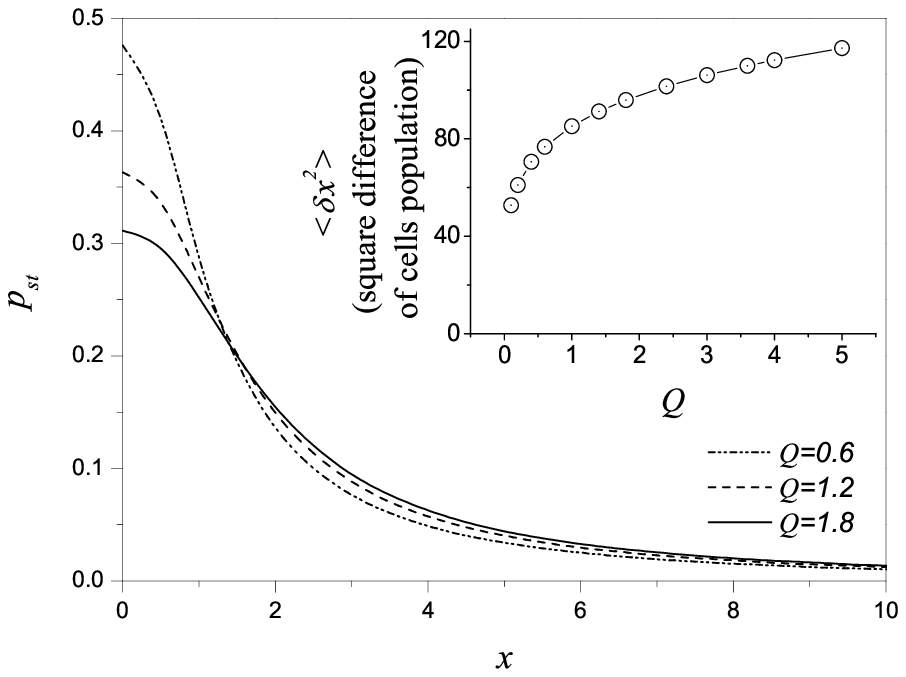';file-properties "XNPEU";}}

\FRAME{ftbpFU}{2.8323in}{2.2096in}{0pt}{\Qcb{STPD under different
correlation at $Q=1.6$ and $D=0.5.$ Inset: the varieties of $\langle
x\rangle $ and $\langle \protect\delta x^{2}\rangle $ with $\protect\lambda %
. $}}{}{graph7.eps}{\special{language "Scientific Word";type
"GRAPHIC";maintain-aspect-ratio TRUE;display "USEDEF";valid_file "F";width
2.8323in;height 2.2096in;depth 0pt;original-width 4.2886in;original-height
3.3382in;cropleft "0";croptop "1";cropright "1";cropbottom "0";filename
'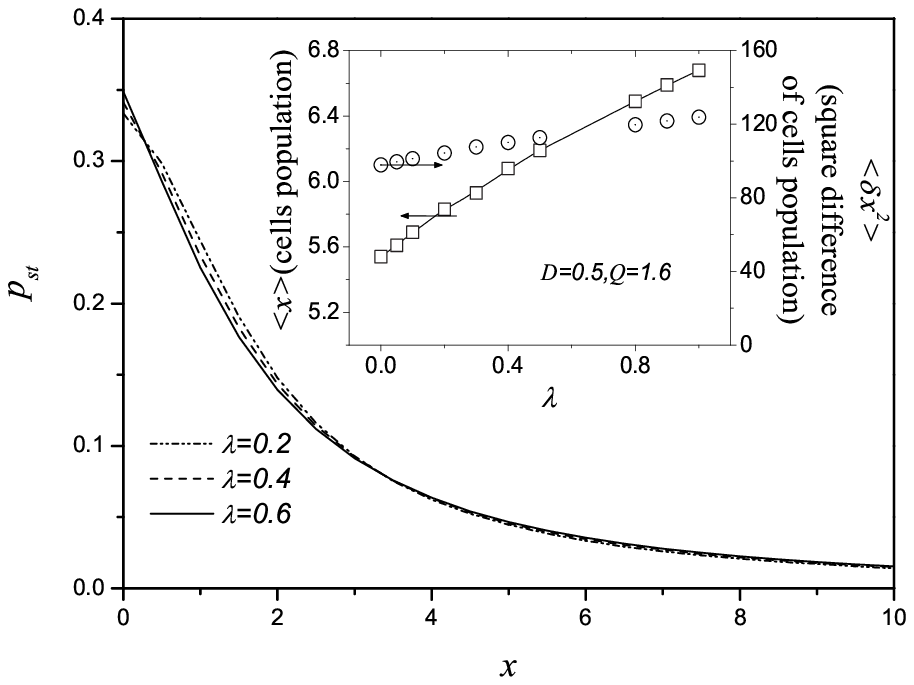';file-properties "XNPEU";}}\FRAME{ftbpFU}{2.8124in}{2.1958in}{0pt%
}{\Qcb{STPD under different multiplicative noise intensities at $\protect%
\lambda =0.0,$ $Q=1.6$. Inset: the varieties of $\langle x\rangle $ and $%
p_{st}(0)$ with $D.$}}{}{graph8.eps}{\special{language "Scientific
Word";type "GRAPHIC";maintain-aspect-ratio TRUE;display "USEDEF";valid_file
"F";width 2.8124in;height 2.1958in;depth 0pt;original-width
4.248in;original-height 3.3105in;cropleft "0";croptop "1";cropright
"1";cropbottom "0";filename '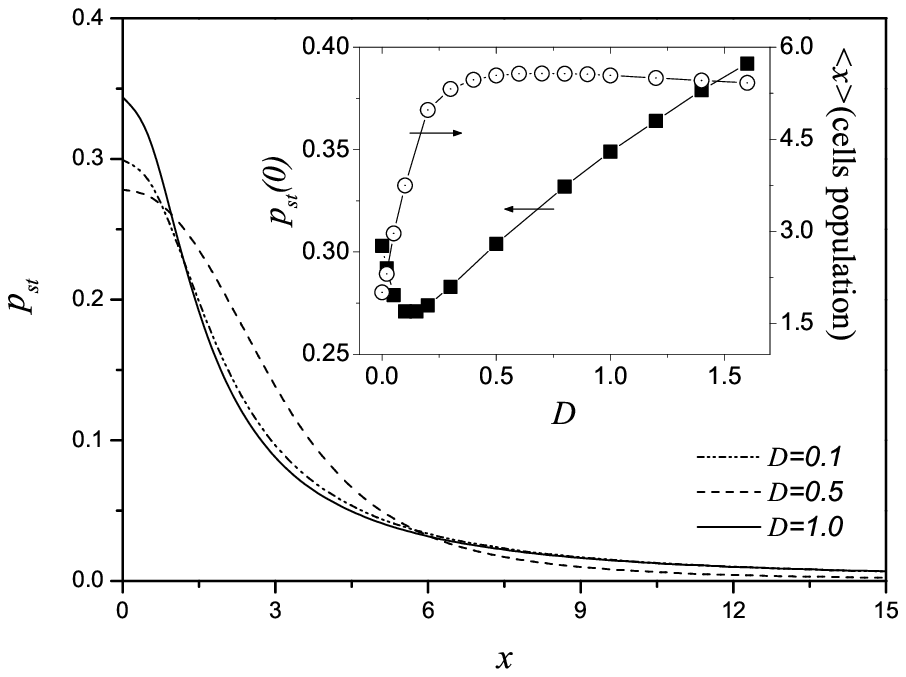';file-properties "XNPEU";}}

\begin{center}
\textbf{IV. DISCUSSIONS}
\end{center}

Tumor growth was largely governed by their intrinsic elements and the
environmental factors like tumor treatment, which was supported through
simulations or experiments [24, 25, 26, 30]. In Ref [25], two kinds of
selection-inducing therapeutic strategies were considered, namely, those
that kill tumor cells and those that block cell mitosis but allowed the cell
to survive for some time. These therapeutic strategies were regarded as the
external multiplicative noise in our model.

There should be a close relationship between the noises and tumor treatment.
Firstly, tumor treatments could cause the fluctuation of the tumor cells
population. The intensity of multiplicative noise is analogous to the
treatment, e.g. the medicinal dose in chemotherapy and the intensity of the
ray in radiotherapy. Secondly, the tumor cells themselves transferred from
one place to the other ceaselessly to survive or encroach. This caused
another fluctuation of the tumor cells. This noise roots in the genetic
instability of the tumor cells themselves, so it is internal noise, i.e.
additive noise. The intensity of additive noise is considered as an
instability of tumor cells population, e.g. genetic instability and cancer
heterogeneity. Thirdly, the correlation between two noises reflected the
adaptation of a tumor to the treatments. If $\lambda =0.0,$ the tumor was
completely unconformable, and if $\lambda =1.0,$ the tumor had better
adjustability. Our model did not refer to a certain therapy, but emphasized
particularly on the response of a tumor to the treatment.

Additive noise induces an unstable tumor growth, but does not cause
extinction. This perhaps owes to the genetic instability and emigration of
tumor cells, which adjust themselves to the environment. Furthermore,
Multiplicative noise induces a phase transition and stochastic resonance.
Firstly, multiplicative noise leads the tumor growth from an uni-stable
state to a bi-stable state. This transition denoted the tumor try to survive
at the price of partial death. In addition, it was confirmed that tumors
grow firstly and then decrease with the enhancement of the treatment. This
stochastic resonance-like characteristic implied an improper therapeutic
intension could not kill the tumors, but lead to their malignant growth. It
was suggested that an enough-strong intensity of therapy was necessary for
tumor treatment. Unfortunately, the normal cells were at the risk of
canceration at the same time. These presumptions implicated a dilemma for
some kinds of therapies like radiotherapy and chemotherapy. Therefore,
provided that some advanced treatments based on repeated applications was
designed, better treatments would be expected.

It was well known that almost all solid cancers were composed of various
genetically and epigenetically more or less different, clonal cell
sub-populations. This cancer heterogeneity referred to the fluctuation of
tumor cells population as well as a better adaptation of the tumor to the
therapies, namely, the additive noise and the correlation of noises. This
dual factors focused on an extreme condition, i.e. strong noise with close
correlation. Tumor cells population, however, increases steadily under this
extreme condition (see inset of Fig.2). This implicated cancer heterogeneity
was disadvantageous to cancer curing. It is one of the challenges in modern
anti-cancer therapy to apply therapeutic strategies to keep up with the
tumor's intrinsic capability to adapt.

The decay case was described as tumor's decreasing under immune
surveillance, which caused a tumor to decay. In this case, additive noise
still played a role of destroyer like in the growth case, and kept tumor
cells far from extinction. Weak multiplicative noise may stabilize the tumor
growth, but strong multiplicative noise accelerated their extinctions.
However, the noises correlation had little effect on the tumor growth law.
These meant that strong-enough treatment and the attempts of different
therapies were still necessary for cancer curing even if there existed
immune systems.

Anyway, tumor treatment includes kinds of complex processes, such as
therapy, immune response, medicine, and so on. Pure therapy just like
chemotherapy or radiotherapy has both negative and positive effects on the
tumors. It was suggested in that more measures should be taken in tumor
treatment to eliminate a tumor completely, such as gene-therapy,
immunotherapy, resection and so on.

\begin{center}
\textbf{V. CONCLUSIONS}
\end{center}

We presumed two kinds of noises in the tumor growth model. One originates
from the tumor treatment, like radiotherapy and chemotherapy; the other
roots in the tumor cells themselves. The previous is multiplicative, and the
latter is additive. On one hand, multiplicative noises induce a phase
transition and stochastic resonance in the tumor growth. An appropriate
intensity of multiplicative noise, e.g. an error treatment, could not kill a
tumor, but activated it. On the other hand, additive noises cause an
unstable growth of\ a tumor, but does not extinct it. Under a weak
multiplicative noise, an additive noise also induces a negative stochastic
resonance. In addition, close correlation between multiplicative and
additive noises, i.e. a better adjustability of a tumor to the environment,
is benefit to the growth of tumors.

\begin{center}
\textbf{ACKNOWLEDGEMENTS}
\end{center}

This work was partially supported by the National Natural Science Foundation
(Grant No. 60471023) and the Natural Science Foundation of Guangdong
Province (Grant No. 031554), P. R. China.

\end{document}